\documentclass[final,12pt,twocolumn]{elsarticle}

\usepackage{amssymb}
\usepackage{amsmath}

\journal{Journal of Systems and Software}

\usepackage{tikz}
\usetikzlibrary{shapes.geometric, arrows, positioning}
\tikzstyle{box} = [rectangle, rounded corners, minimum width=3cm, minimum height=1cm,text centered, draw=black, fill=gray!30]
\tikzstyle{arrow} = [thick,->,>=stealth]
\usepackage{booktabs}
\usepackage{tabularx}
\PassOptionsToPackage{hyphens}{url}
\usepackage{url}

\begin{document}

\begin{frontmatter}



\title{Challenges and Opportunities: Implementing Diversity and Inclusion in Software Engineering University Level Education in Finland}


\author{Sonja M. Hyrynsalmi} 

\affiliation{organization={Department of Software Engineering, LUT University},
            addressline={Mukkulankatu 19}, 
            city={Lahti},
            postcode={15210},
            country={Finland}}


\begin{abstract}
Considerable efforts have been made at the high school level to encourage girls to pursue software engineering careers and raise awareness about diversity within the field. Similarly, software companies have become more active in diversity and inclusion (D\&I) topics, aiming to create more inclusive work environments. However, the way diversity and inclusion are approached inside software engineering university education remains less clear. This study investigates the current state of D\&I in software engineering education and faculties in Finland. An online survey (N=30) was conducted among Finnish software engineering university teachers to investigate which approaches and case examples of D\&I are most commonly used by software engineering teachers in Finland. In addition, it was researched how software engineering teachers perceive the importance of D\&I in their courses. As a result of the quantitative and thematic analysis, a framework to identify attitudes, approaches, challenges and pedagogical strategies when implementing D\&I themes in software engineering education is presented. This framework also offers a process for integrating D\&I themes for the curriculum or at the faculty level. The findings of this study emphasize that there is a continuing need for diverse-aware education and training. The results underline the responsibility of universities to ensure that future professionals are equipped with the necessary skills and knowledge to promote D\&I in the field of software engineering.

\end{abstract}


\begin{highlights}
    \item Diversity and inclusion (D\&I) are growing in importance in the software industry
    \item A study (n=30) of Finnish software engineering (SE) university-level educators is conducted
    \item Some educators are hesitated to add D\&I topics; yet, the view is neutral or positive
    \item A framework summarizing attitudes, approaches, challenges, pedagogical strategies and the D\&I integration process for SE is formulated
\end{highlights}


\end{frontmatter}



\maketitle

\section{Introduction}
Approaching the topic of diversity and inclusion (D\&I) is not always easy, as they are broad themes with different viewpoints towards them~\cite{williams1998demography}. On the one hand, at the general higher abstraction level, 'diversity' refers to representing individuals with varying backgrounds or perspectives. The most commonly recognized forms of diversity are related to \textit{social diversity} such as gender, age, and ethnicity. On the other hand, 'inclusion' means that individuals with diverse backgrounds feel welcomed, have equal access to participate, and are viewed as equal group members. Given the significant impact, challenges, and opportunities related to diversity and inclusion, this theme has piqued the interest of researchers from various fields.~\cite{jehn1999differences,roberson2006disentangling,lirio2008inclusion,daya2014diversity}

In the field of software engineering, it can be stated that certain aspects of diversity have been researched more than others. The research around gender, especially women, in software engineering has been under study for decades~\cite{margolis2002unlocking}. The study by Rodríguez-Pérez et al.~\cite{rodriguez2021perceived}, for instance, indicates that gender bias is the most commonly studied aspect of diversity in software engineering. However, gender is only one aspect related to social diversity. Nowadays, diversity and inclusion topics in software engineering research also include a more diverse set of research aspects.~\cite{rodriguez2021perceived} For example, there has been increased research on topics such as ethnicity, nationality, age, and their intersections. This may be attributed to the growing popularity of the intersectional approach in software engineering research. Intersectionality~\cite{crenshaw1989demarginalizing, mccall2005complexity} is a term and framework emphasizing the importance of different experiences based on various backgrounds. For example, if experiences are studied only from the viewpoints of white, middle-class women, the results may not be as applicable or relatable for women of colour or those working in low-wage sectors. 

Diversity and inclusion are important to the software industry in many ways. Nurturing team diversity and creating an inclusive work environment can positively impact software engineering practices, design, and requirement engineering~\cite{menezes2018diversity}. It has been researched that diverse and inclusive teams perform better than homogeneous teams~\cite{altiner2018approach}, especially when they are encouraged to recognize and appreciate the potential and value of their team's diversity~\cite{homan2007bridging}. Better team performance not only benefits one team or group's performance but also impacts the entire enterprise level and the company's competitiveness~\cite{page2007power}.

To develop effective strategies for promoting diversity and inclusion in education and faculties, it is crucial to understand software engineering educators' barriers, attitudes and perceptions towards D\&I and how these different viewpoints currently translate into educational practices. This study aims to investigate the current state of  D\&I  in software engineering education and faculties in Finland with the following two research questions: 
\begin{description}
\item[RQ1] Which specific approaches and case examples of  D\&I  are most commonly used by software engineering teachers in Finland?
\item[RQ2] How do software engineering teachers in Finnish universities perceive the importance of  D\&I  in their courses, and what challenges do they face in incorporating these topics into their teaching?
\end{description}

To address the research questions, an online survey was conducted among software engineering university teachers and professors from six Finnish universities offering software engineering study programs. A total of 30 usable responses were received. The study examined the prevalent approaches to diversity and inclusion in software engineering university-level teaching, the reasons for teachers' use or non-use of these approaches, and the challenges or difficulties they encounter when attempting to incorporate more diversity and inclusion in their courses.

This article extends the previous work~\cite{Hyrynsalmi2023} by extending the analysis. The results are enhanced by incorporating new categorization of the cases when adding more D\&I aspects to software engineering education. The extended work also provides a new framework for identifying challenges, attitudes, approaches, and cases in adding D\&I to software engineering university-level education and steps for the integration process for D\&I in the faculties. The background section has also been updated to discuss more with the results to get the state-of-art about the D\&I in software engineering education. 

The remaining of this study is structured as follows. Section~\ref{sec:background} discusses related studies on teaching diversity and inclusion in universities and the industry. Section~\ref{sec:method} goes through the used research approach, and the results are reported in Section~\ref{sec:results} and the framework presented in~\ref{sec:framework}. Finally, the findings are discussed in Section~\ref{sec:discussion}, and the study is closed in Section~\ref{sec:conclusions}.

The survey design and quantitative data are available in the dataset\footnote{\url{https://doi.org/10.5281/zenodo.12791197}} accompanying this publication. 

\section{Background}\label{sec:background}
The study of diversity and inclusion can be considered challenging in particular because it involves multiple layers and classifications, such as diversity of values (i.e. attitudes and beliefs), diversity of information (i.e. knowledge and experience), and social diversity (i.e. ethnicity, age and gender) studied in general. Therefore, this challenges the analysis, methods, and generalizability of results.~\cite{jehn1999differences}

As a result, the research in D\&I in software engineering poses a significant challenge regarding the generalizability of results. When studying software teams, researchers need to consider various dimensions of the respondents, such as cultural background, age, gender, past experience, values, and more. Therefore, it is crucial to exercise special sensitivity when drawing theories or conclusions about the topic.~\cite{menezes2018diversity}

The upcoming subsections will examine prior research on teaching diversity and inclusion, as well as the industry's and software engineering teaching approaches to the topic.

\subsection{Teaching diversity and Inclusion in universities}

There are various ways to incorporate D\&I approaches into university teaching~\cite{meletiadou2022handbook}. These may include discussing the topic, inviting guest speakers, incorporating journaling and self-reflection exercises, delivering lectures, assigning readings, and requiring formal papers~\cite{boysen2011diversity}. 

Embedding diversity topics into the curriculum and teaching is often a slow process and may face resistance. However, D\&I topics are becoming more critical in the academic world, and in that way, they are getting implemented in the curriculums of various fields.~\cite{fuentes2021rethinking, wolbring2021equity} Therefore, researching best practices and experiences from other university fields can be helpful. For example, during a discussion on adding more teaching on cultural diversity and embedding it into the medical curriculum, twelve tips were shared~\cite{dogra2009twelve}. From these tips, 1-2 were focused on institutional policies, 3-8 on the curriculum, 9-10 on faculty development, and 11-12 on the assessments: 
\begin{enumerate}
    \item Design a diversity and human rights education institutional policy,
    \item Create a safe learning environment,
    \item Develop clear and achievable learning outcomes,
    \item Develop content focused on the diversity of human experience, 
    \item Raise awareness of students’ own biases and prejudices,
    \item Integrate cultural diversity across the entire curriculum, 
    \item Make diversity patient-centred,
    \item Teach outside the classroom and hospital setting, 
    \item Form multi-disciplinary teams of educators,
    \item Make training of faculty compulsory,
    \item Develop a clear and comprehensive assessment of policies, delivery and learned outcomes of cultural diversity education, and
    \item Map what others are doing and challenge yourself as a role model.
\end{enumerate}

There are several pedagogical strategies to add more diverse approaches to the curriculum. To foster students’ awareness, knowledge, skills, and capacity towards D\&I, there have been identified five potential approaches: (1) experiential learning, (2) activation of students’ prior knowledge, (3) repositioning the role of the instructor, (4) community-based learning, and (5) reflection. ~\cite{hartwell2017breaking} Inside of these approaches could fit some of the most popular tools or methods to use in education, such as 'intergroup dialogue'. For example, Intergroup dialogue is used in social justice education. In that students from at least two social identity groups agree to meet and have a conversation to build relationships and dialogue across, for example, cultural and power differences.~\cite{gurin2013dialogue, nagda2007intergroup}

When adding more D\&I approaches to the curriculum, it is important to ensure everyone feels safe, has equal access to learning, and feels valued and supported in their pursuit of knowledge~\cite{kayingo2017health, laverty2022student}. When researching the students' perspectives on inclusive teaching, three themes were identified as key factors contributing to inclusive education: persona, practice, and partnership.~\cite{laverty2022student}

The development of D\&I concepts can be challenging as they are impacted by cultural, social and organizational aspects. Furthermore, organisations can use the same concept or method in totally different ways.~\cite{may2010developing, awang2019strategizing} But because of the complexity of the topic, concepts and frameworks developed around D\&I usually have multiple steps to ensure that everyone has the possibility to participate and the process is clear for participants. For example, Pless \& Maak's~\cite{pless2004building} framework of inclusion has four different steps: The first phase focuses on raising awareness, building understanding, and encouraging reflection. The second phase involves developing a vision of inclusion to guide the change. The third phase reconsiders key management concepts and principles. The fourth, action-oriented phase implements an integrated Human Relations Management (HRM) system, translating principles into observable behaviour and fostering inclusive practices through development and recognition.

\subsection{Diversity and inclusion in software engineering}

One approach to how people see that they belong in some field is how well presented they feel that people similar to them are represented in the field. Gender has been one of the primary aspects of diversity and inclusion research in software engineering, as there is still a significant gender gap in the software engineering field. Gender-focused research has also made visible how important factor gender diversity is in software development.~\cite{menezes2018diversity, rodriguez2021perceived} 

As research on diversity and inclusion in software engineering continues to advance, multiple approaches (often including gender) are being used to study the topic. One such approach is age, which explores the ongoing public discourse on age, age discrimination, and software development~\cite{baltes202040}. Furthermore, a holistic or intersectional approach can enable software engineering researchers to delve deeper into the topic of D\&I, developing specific actions based on research findings~\cite{sanchez2021framework}. 

Teaching diversity and inclusion in software engineering courses is different than, for example, making sure that pictures in the course slides are diverse or that the course materials or lecture hall are accessible. In recent years, there has been a need for more visibility for underrepresented groups in computer science and software engineering education~\cite{silveira2019systematic}. For example, LGBTQIA+ students feel more secure and welcomed in fields such as the social sciences and humanities, and that gives a sign for the field of software engineering that we still have work done around equity, diversity, inclusion, and belonging~\cite{de2023lgbtqia+}. However, Gen AI is transforming the teaching of diversity and inclusion as software engineering, influencing both educational materials and teaching methods~\cite{hyrynsalmi2023diversity, hyrynsalmi2024bridging}

One potential way to integrate inclusive teaching methods and experiential training into software engineering education is by incorporating them into project courses, mainly through intensive Capstone courses. Capstone courses are a popular model of project courses that can provide opportunities to introduce themes such as communication, leadership, and empathy skills in teamwork and retrospectives.~\cite{janzen2018reflection} These skills are usually referred to as 'soft skills', and their importance in software engineering has grown recently. However, they are still not fully recognized in higher education curriculum~\cite{gonzalez2011teaching, matturro2019systematic} as it is also the situation for diversity and inclusion topics. 

\subsection{Diversity and inclusion actions taken by software companies}

There is now a wide range of different D\&I training offered to software companies. These trainings usually contain hands-on exercises, such as experiential role-playing~\cite{pendry2007diversity}. In organizational settings, the goal of diversity and inclusion training is to promote a more inclusive and diverse workplace climate. These training sessions also help organizations pay attention to more inclusive recruitment practices and familiarize human service professionals with culturally based beliefs.~\cite{devine2021diversity}

For example, software companies have hired diversity and inclusion professionals to assist with recruitment advertisements to attract more women to apply for their open positions~\cite{Tietoevry}. As the number of diversity and inclusion training options grows, it is essential to recognize their limitations and potential pitfalls. One of the most significant limitations of diversity training is the mismatch between its goals and evaluated efficacy. Therefore, it is crucial to establish reasonable goals for diversity and inclusion initiatives and focus on long-lasting efforts.~\cite{devine2021diversity}

When software professionals enter the software industry, they have worked through the education path from kindergarten, middle and high school, and university. Therefore, the software industry, as before the education levels~\cite{margolis2002unlocking}, plays an important role in keeping underrepresented groups, such as women or people from LGBTQIA+ groups, in the field. That requires that the field is safe, people feel valued, and the work environment is professional and motivational~\cite{hyrynsalmi2019motivates, de2023lgbtqia+, hyrynsalmi2019software}. Still, even recent studies show that there are, for example,  still challenges for women in the field of software engineering~\cite{trinkenreich2022empirical, hyrynsalmi2019motivates, hyrynsalmi2020meaningfulness}. Therefore, it is also important for the industry to support and follow diversity and inclusion research and participate in it by itself.  

\section{Research Methods}\label{sec:method} 

The data for this study was collected through an exploratory online survey conducted in September-October 2021. The survey was distributed to 140 Finnish software engineering university teachers, including university teachers, lecturers, docents, PhD researchers, professors, etc. Out of the 140, 30 individuals responded to the survey after two reminder rounds, resulting in a response rate of 21.4\%. The survey consisted of 11 questions, two of which were descriptive (regarding gender and years of teaching experience), four were structured, and five were open-ended. The questionnaire was opened by 59 different individuals, resulting in a 42.1\% open rate. All those 30 respondents who started the questionnaire answered all questions.

The sampling strategy, survey design, testing, and analysis process are presented in the following subsections.

\subsection{Sampling strategy} 

To identify the most suitable respondents, the study guidance for the academic year 2021--2022 was reviewed in all six Finnish universities that offer software engineering as a major. From the study guidances, software engineering bachelor's and master's degree courses were identified, and the names and emails of the teachers responsible for them were collected. 

In this phase of the research project, the focus was only on the field of software engineering to obtain a state-of-the-art understanding of diversity and inclusion topics in software engineering education. Therefore, courses and teachers from other fields, such as mathematics, included in the study programs were excluded. Teaching assistants were also excluded from this phase because 1) there was not enough information or contact details available for all of the teaching assistants at every university, and 2) teaching assistants usually are not in charge of the content of the lectures or demonstrations and practice sessions in Finnish universities. 

In the sampling strategy phase, it was noted that courses offering diversity and inclusion approaches for students could also be provided outside the software engineering department. One example could be a course on communication skills and professional behaviour provided at least in one university by the Language Center. This is one of the limitations that should be considered in future research designs.

\subsection{Survey design and testing} 
In this research, the aim was to gain an understanding of how diversity and inclusion are addressed in university-level software engineering teaching. An anonymous online questionnaire was created to investigate that. In the recruitment email for identified respondents, those who thought that they did not address any diversity and inclusion themes in their teaching were specifically encouraged to respond. This was done to avoid generalizability threats, getting only answers from those who would be already interested in the topic and who would presumably be more likely to pay attention to diversity and inclusion themes in their teaching. Therefore, each time the invitation or reminder email was sent to the potential respondents, the need for responses from those participants unfamiliar with the theme was emphasized to avoid generalizability threats.

In this study it was emphasized that this research study does not focus on diversity and inclusion from the perspective of how students can participate in the courses. Attention was paid to the message sent to the potential respondents, which made it clear that this research was not asking about designing courses for people with disabilities or learning challenges, such as dyslexia. While accessibility of courses is an important topic, most universities have their strategies for addressing this issue. Additionally, these strategies are usually supported by country-specific legislation on participation and equal treatment.

For the online survey testing phase, the walk-through method~\cite{light2018walkthrough} was employed to gauge how well the questions were understood and answered by someone not well-versed in diversity and inclusion topics. One university teacher was chosen for the testing and to modify the questions in two brainstorming sessions before testing. After the testing phase, further modifications were made to the questions to ensure that impactful responses were obtained.

One example of how impactful answers were ensured during the testing involved a checkbox option for different approaches being used in the teaching. After testing the questionnaire, it was noticed that the results for this question in the first draft did not provide sufficient information. Moreover, it only gave a list of approaches used without further detail. Therefore, the first question was modified to ask respondents to score the approaches they use, with the most used approach receiving a score of 1, the second most used receiving a score of 2, and so on. It was mandatory for the respondents to choose at least one option, and an option was provided for "I don't use any examples about diversity and inclusion in my course(s)."

During the data collection, two feedback emails were received from university teachers who had been contacted for the survey. One feedback stated that it was difficult to answer the questions because the whole topic was unfamiliar. The other feedback expressed a desire for more personal interviews to ensure an understanding of the examples. Both of these responses were from university teachers with more than 20 years of teaching experience. At the beginning of the research project it was acknowledged that the terminology could be challenging. Therefore, at the beginning of the survey and the participation invitation letter, terms such as diversity and inclusion were explained in the following way: \textit{"Diversity means the representation of people with different background, from different cultures, ethnical group, age groups, different genders. Inclusion means that people with different backgrounds are welcomed and have access to participate and feel part of the group. In this short questionnaire, I use "D\&I" for Diversity \& Inclusion and "SE" for Software Engineering."}

In this research, the phrase "Diversity and Inclusion" (D\&I) is used as it was used in the survey. However, in the results it is also broadened to the topics around EDI (equity, diversity, and inclusion) or (DEI) (diversity, equity, and inclusion).

\subsection{Analyze method}

The survey consisted of 11 questions, of which four were structured, and five were open-ended. While the open-ended responses varied in topics, some respondents touched on similar themes in multiple questions and reporting answers directly from every question would not give a broad enough perspective from the results. To identify the main themes across all the responses, a thematic analysis~\cite{Braun2008} was conducted for the results of the five open-ended questions. Some of the questions also included some support questions for the respondents.


Thematic analysis is a qualitative method often used for identifying and categorizing patterns or themes from data~\cite{Braun2008}. Thematic analysis was chosen for this study due to its suitability for exploring complex and multifaceted phenomena like diversity and inclusion in software engineering education. Furthermore, it allows an in-depth examination of data without requiring predefined hypotheses, which is essential for capturing the educators' varied experiences and perspectives. In addition, thematic analysis is a widely accepted and well-established method in software engineering.~\cite{Badreddin2013, Cruzes2011a, Cruzes2011b}

The coding and analysis phase was conducted following the guidelines introduced by Braun \& Clarke~\cite{Braun2008}. The first phase began with familiarizing with the data, during which some notes and ideas were marked for coding. There was no need for transcription of verbal data, as the data was collected from an online survey's open-ended questions. Then, the initial data-driven codes were generated. Following this, the codes were organized into potential themes and in the reviewing phase, themes were redefined and merged one more time to create a thematic map. Finally, the themes were reviewed, and the first version of the visual framework was conducted.

The results of the open-ended questions identified four cross-cutting themes: approaches used, reasons for adding or not adding diversity and inclusion topics to teaching, challenges faced in adding more diversity and inclusion topics, and overall satisfaction with the faculty's actions toward the topic. The qualitative data totalled 11,576 characters with spaces and 72 paragraphs.

\subsection{Threats to validity}

Ensuring the validity of the findings is a critical aspect of this study, as the D\&I is a sensitive human topic to research. Therefore, the threats to validity were handled in terms of internal, construct, conclusion and external validity~\cite{Wohlin2012}. 

Internal validity—A common challenge in online surveys is that they do not necessarily attract respondents who are not interested in the topic. For that reason, the participation invitation letter strongly emphasized that those people who are not interested in D\&I topics would be needed to answer.

Construct validity -- The terminology around D\&I, especially what it could mean in software engineering, could cause misunderstandings. To ensure that all the participants could share a similar approach to the topic, definitions and clarifications were offered at the start of the survey.

Conclusion validity -- The Finnish focus and the small sample of the respondents could be seen as threats to conclusion validity. However, the survey has been designed from the beginning to be conducted for an international audience, and it does not, for example, have any questions about Finnish-specific legislation or support systems.  

External validity -- As the Finnish focus also goes to the external validity threats, so goes the challenge around survey design regarding the academic position. As Finland is a small country and academic software engineering circles are well-connected, the question around academic title was left out to lower the barrier to answer. However, when the survey is used, and the framework is tested for a bigger audience, adding the question about the academic position will become more important. 

By addressing these potential threats to validity, this study sought to ensure reliable and rigorous findings regarding integrating diversity and inclusion themes in software engineering education.

\section{Results}\label{sec:results} 
In the descriptive questions, respondents were asked how many years they had been teaching and their gender (Table~\ref{exp}). The respondents were relatively experienced, as 50\% reported having ten years or more of teaching experience in software engineering courses at the university level. 

Answering the gender question was optional, and respondents could also provide their own definitions in the blank box. Sixteen respondents identified as men, and seven respondents identified as women.
\begin{table}[t]
\centering
\footnotesize
\begin{tabular}{l rr}
\toprule
\textbf{Category} & \textbf{n} & \textbf{Percent} \\ 
\midrule
\textbf{Teaching experience in years} & & \\
\hspace{0.5cm} $<$ 5 & 5 & 16.7\% \\
\hspace{0.5cm} 5-10 & 10 & 33.3\% \\
\hspace{0.5cm} 10-20 & 11 & 36.7\% \\
\hspace{0.5cm} 20+ & 4 & 13.3\% \\
\midrule
\textbf{Gender} & & \\
\hspace{0.5cm} Men & 16 & 69.6\% \\
\hspace{0.5cm} Women & 7 & 30.4\% \\
\bottomrule
\end{tabular}
\caption{Demographic information of the respondents.}\label{exp}
\end{table}
The thematic analysis was conducted on the responses to five open-ended questions, as in many questions, respondents highlighted issues that could have been reported under some other questions. The main idea was to find cross-cutting themes from the data. The open-ended questions were: 1) Why do you use or why don’t you use diversity \& inclusion examples in your course?, 2) What other examples/approaches do you use in your teaching about diversity \& inclusion?, 3) Could you provide some examples of how and where you use diversity \& inclusion examples in your teaching?, 4) Have there been some challenges in adding more info about D\&I in SE teaching and 5) Any other comments, thoughts about the topic etc.?

There were 70 unique responses by 30 respondents for these five questions, and from those responses, 7 were just "nothing to add to this, nothing more comes to my mind just right now, this was an awaking survey, thank you" kind of responses. From the 63 valid responses, 40 were labelled as positive, ten were neutral, 8 were reluctant answers, and five answers were categorized as miscellaneous/unclear. Some responses were also labelled "challenges" or "Pedagogical strategies". In the following subsections, we will present the results of the thematic analysis combined with the results of the quantitative questions in the subsection "Diversity and inclusion approaches in software engineering education in Finnish universities" and "Overall satisfaction to the faculty's diversity and inclusion actions".

\subsection{Overall satisfaction with the faculty's diversity and inclusion actions}

Table~\ref{perceptions} reports respondents' perceptions on the importance of raising diversity and inclusion topics in the quantitative questions: "How important do you see that diversity and inclusion issues are approached in software engineering teaching?", "How important do you see that diversity and inclusion issues are raised up in your own software engineering faculty/team/school?" and "How satisfied are you on how diversity and inclusion issues are approached in your software engineering faculty/team/school?". The results show that there was little difference in importance between teaching and faculty actions. However, respondents expressed dissatisfaction with how diversity and inclusion issues were approached in their faculty/team/school, with an average score of 6.3 out of 10.  
\begin{table*}[t]
\footnotesize
\begin{tabularx}{\textwidth}{p{6cm} XXXX XXX}
\toprule
\textbf{Question}	&	\textbf{Mini\-mun value}	&	\textbf{Maxi\-mum value}	&	\textbf{Average}	&	\textbf{Median}	&	\textbf{Sum}	&	\textbf{Standard deviation}	\\
\midrule
How important do you see that diversity and inclusion issues are approached in software engineering teaching?	&	3	&	10	&	7,6	&	8	&	228	&	2	\\
\hline
How important do you see that diversity and inclusion issues are raised up in your own software engineering faculty/team/school?	&	3	&	10	&	7,5	&	8	&	226	&	2,3	\\
\hline
How satisfied are you on how diversity and inclusion issues are approached in your software engineering faculty/team/school?	&	2	&	10	&	6,3	&	6,5	&	190	&	2,2	\\
\bottomrule
\end{tabularx}
\caption{Ratings of Importance and Satisfaction with Diversity and Inclusion Approaches in Software Engineering Education and Faculty}\label{perceptions}
\end{table*}
Nevertheless, as revealed in the thematic analysis of the open-ended responses, respondents generally viewed the academic field as diverse and multicultural, and they were satisfied with the actions taken by their faculty. The focus on internationalization and the number of international staff members was especially seen positively. Respondents felt that they were satisfied with the actions in this area:
\begin{quote}
   \textit{ "I think at the faculty level, these topics are well presented. We do have more and more people from different cultural backgrounds."}
\end{quote}
Among the number of international staff members, many respondents also provided examples of gender-diverse faculty actions and the aims and impact of those projects. There were several projects and successful best practices both at the department and the university level:
\begin{quote}
    \textit{"We have a departmental project where we try to attract more women to IT and support and give visibility to those we already have."}
\end{quote}
An important finding of the study was the presence of trust issues and confrontation between individuals who prioritize diversity and inclusion topics and those who do not. This observation suggests a potential challenge in promoting and implementing diversity and inclusion initiatives:
\begin{quote}
   \textit{ "I think challenges come from negative attitudes and lack of information, though not from students but from whom could be called 'older colleagues'."}
     \end{quote}
Some respondents expressed a slightly pessimistic view that the primary obstacle to addressing diversity and inclusion issues is the reluctance of some teachers to engage with or incorporate relevant approaches into their courses. 
    \begin{quote}
    \textit{"I don't think the students are the problem; more, I think that some teachers are not so willing to really discuss inclusion topics - not in the faculty and not in their courses."}
\end{quote}

\subsection{Attitudes towards Diversity and Inclusion Topics in the Teaching of Software Engineering}

In the thematic analysis conducted, three distinct approaches were identified as to why diversity and inclusion topics should or should not be included in software engineering teaching: \textit{positive}, \textit{neutral}, and \textit{reluctant} (Table \ref{attitudes}).

\subsubsection{\textbf{Positive responses}}

\begin{table*}[h!]
\footnotesize
\centering
\begin{tabular}{| m{5cm} | m{5cm} | m{5cm} |}
\hline
\textbf{Positive responses (N = 40)} & \textbf{Neutral responses (N = 10)} & \textbf{Reluctant responses (N = 8)} \\ \hline
Values team diversity & Does not see D\&I themes as different than other approaches & Does not see any reason to add D\&I topics to SE education \\ 
See project courses as a good way to add D\&I themes to SE education & Focus on technical approaches rather than human approaches & Suspicious about the advocacy behind D\&I themes \\ 
Highlights human-centered focus & Uses D\&I viewpoints only if really clearly suitable for the course theme & Believes that D\&I themes in SE could be more harmful than helpful \\ 
See impact on the future & & \\ \hline
\end{tabular}
\caption{Responses regarding D\&I themes in SE education}\label{attitudes}
\end{table*}

Most of the comments on the topic were viewed as positive in the thematic analysis. Positive responses were categorized and identified to value team diversity, see project courses as a good way to add D\&I themes in SE education, have a human-centred focus in their teaching or courses and believe that D\&I is important because it has an impact on the future.

Positive responses emphasized the importance of diversity of skills. These teachers saw teamwork skills as crucial for future software developers to learn. They also had a wide range of tools or processes for approaching teamwork skills in their courses.

\begin{quote}
\textit{"I often focus on team diversity, which is super important for great ideas and divergent thinking. Diversity brings diverse views and knowledge. This also applies to diversity in problem-solving and innovations. I do have specific methods to ensure diversity and team composition."}
\end{quote}

As mentioned, the project courses were seen as an ideal place to discuss diversity and inclusion topics in software engineering, and many teachers who were responsible for software project courses responded to this survey. They highlighted that there could be different opinions among the teams and different roles in the companies: 

\begin{quote}
\textit{"I emphasize that most software designs should start from people's needs, and people can be different from the designers. In one course, I discuss diversity and diversity of skills in design teams.}"
\end{quote}

Among project courses, software design courses were also seen as a good starting point for implementing more diversity and inclusion awareness in the curriculum. Teachers also regularly used the opportunity that they would not allow students to pick their own team; instead that, teachers would form the teams by themselves, or there would be some kind of process for creating a diverse team: 

\begin{quote}
\textit{"I teach a course in HCD (Human-Centered Design) where the main idea is what IT engineers need to know about human (neuro)psychology. Diverse teams are a must (I form the teams), and accessibility of all kinds is very much in focus."}
\end{quote}

Overall, teachers emphasized the importance of diversity and inclusion in the viewpoint of the future of software development and products. The focus was especially on that they were teaching future software engineers and that they should know about D\&I topics in their work:

\begin{quote}
\textit{"These topics are vital to software engineering professionals and how these professionals will shape the world in the future."}
\end{quote}

Teachers pointed out that this theme is broader than just web application development. Software engineering has a cross-cutting impact on our society, and teachers believe that software engineers have an essential role in creating more sustainable societies in the future. Teachers felt that it was their responsibility to ensure that students graduate with an understanding of their influence and the decisions made in software development.

\begin{quote}
\textit{"I think that we create better products and services with diversity; I also think that we create a better, socially sustainable world for everybody by teaching our students socially responsible ways of working. It is our responsibility as teachers of future professionals to do that. The course topics are not related to diversity and inclusion."}
\end{quote}

From case examples, biased data was seen as really easy to teach, and overall, students seemed to like real-world cases that might impact their own lives. Teachers also enjoyed examples of biased data themselves. They felt safe adding these kinds of diversity and inclusion approaches, as it was easy to show the connection to the software development topic.

\begin{quote}
\textit{"I use mostly biased data examples. I found them interesting for myself. They are also easy to understand in software development content."}
\end{quote}

\subsubsection{\textbf{Neutral responses}}
    
In the thematic analysis, neutral responses were categorized and identified in a way that does not see D\&I themes as different from other approaches. They say to focus on technical approaches rather than human approaches and use D\&I viewpoints only if they are really clearly suitable for the course theme.

Most of the neutral responses were identified in a way that indicated that diversity and inclusion themes were not deliberately added to the course or were added without much forethought. Some teachers mentioned focusing solely on technical rather than human approaches. However, these responses did not necessarily oppose adding more diversity and inclusion approaches to the curriculum: 
    \begin{quote}
    \textit{"No, I have not added D\&I by purpose. In general, I am more interested in technology than people."}
    \end{quote}
Other neutral responses indicated that certain diversity and inclusion themes naturally arise in the content of the course. For example, in a web development course, accessibility themes may be a natural curriculum component. In these cases, teachers did not feel the need to add diversity and inclusion themes deliberately. One respondent explained: 
    \begin{quote}
    \textit{"My teaching is mainly on paradigms and technologies, not that much on software engineering processes or development work/teams. Many D\&I aspects are therefore not relevant in my opinion"}
    \end{quote}
Finally, some teachers mentioned that they had added some mentions of diversity and inclusion topics to their lectures but had not fully explored the possibilities for incorporating more diverse and inclusion-aware approaches into their teaching and did not see D\&I themes as different from other approaches. For example, one respondent noted:  
    \begin{quote}
   \textit{ "I have some lecture material that gives examples of the achievements of women in programming and computer engineering (for example, Ada Lovelace and Margaret Hamilton of NASA). Other than that, my teaching has not touched these issues."}
\end{quote}

\subsubsection{\textbf{Reluctant Responses}} 

Reluctant responses were a minority among the open-ended responses. This could be attributed to the fact that most respondents were already familiar with diversity and inclusion themes and had incorporated them into their teaching. In the thematic analysis, reluctant responses were categorized and identified in that way that they do not see any reason to add D\&I topics to SE education, they are suspicious about the advocacy behind D\&I themes, and they believe that D\&I themes in SE could even be more harmful than helpful in some cases.

Respondents with reluctant responses did not favour having diversity and inclusion topics forced into the curriculum. Some even questioned whether these themes were politically biased and had no place in software engineering teaching, as in this response:
\begin{quote}
\textit{"While I believe it is important to treat all students equally and make courses accessible to those from diverse backgrounds, I am not in favour of including diversity and inclusion examples in course materials. Reviewing course materials to ensure they do not offend anyone or enforce negative stereotypes is important. Incorporating diversity and inclusion examples in software engineering materials can be seen as an attempt to impose a political agenda on one's teaching, which I would like to avoid."}
\end{quote}
However, some respondents tried to balance their reluctant stance with a more neutral response. As the survey was conducted anonymously online, it was not possible to determine what reluctant responses were really opposed to:
\begin{quote}
\textit{"People solve problems differently, and having some degree of diversity while solving problems and working in teams is a wise approach. However, diversity should not be enforced, as it may negatively affect performance."}
\end{quote}
In some cases, respondents appeared to hold negative and reluctant attitudes toward the topic but did not express them explicitly in their responses. It could be understood that they were trying to verbalize that sometimes D\&I themes in SE could be more harmful than helpful:
\begin{quote}
\textit{"Some diversity examples may have good intentions but could be overwhelming for individuals from minority groups."}
\end{quote}
Some teachers stated that they wished to teach only technical approaches, and they did not see any place for D\&I themes in the highly technical courses. However, as one of the respondents was describing, they usually understood what diversity and inclusion in teaching could mean and have still done some actions, such as making materials suitable for diverse audiences:
\begin{quote}
\textit{"Except for making the course materials accessible and suitable for a diverse audience, I do not see the need or importance of using diversity examples in software engineering courses of a highly technical nature."}
\end{quote}

\subsection{Diversity and inclusion approaches in software engineering education in Finnish universities} 
\begin{table*}[!bth]
\scriptsize
\begin{tabularx}{\textwidth}{p{2,5cm} rrrrr rrrrr l}
\toprule
\textbf{Approach / Case example}	&	\textbf{1}	&	\textbf{2}	&	\textbf{3}	&	\textbf{4}	&	\textbf{5}	&	\textbf{6}	&	\textbf{7}	&	\textbf{8}	&	\textbf{9}	&	\textbf{10}	&	\textbf{n}	\\
\midrule
Accessibility	&	58,8 \%	&	23,5 \%	&	0 \%	&	11,8 \%	&	0 \%	&	0 \%	&	0 \%	&	0 \%	&	5,9 \%	&	0 \%	&	17	\\
\midrule
Diverse teams	&	33,3 \%	&	26,70 \%	&	6,60 \%	&	6,70 \%	&	20,00 \%	&	0 \%	&	0 \%	&	6,7 \%	&	0 \%	&	0 \%	&	15	\\
\midrule
Un\-der\-rep\-re\-sen\-tation of women in Software Engineering	&	14,3 \%	&	28,5 \%	&	14,3 \%	&	14,3 \%	&	14,3 \%	&	14,3 \%	&	0 \%	&	0 \%	&	0 \%	&	0 \%	&	7	\\
\midrule
Biased data 	&	18,2 \%	&	18,2 \%	&	18,2 \%	&	0 \%	&	27,2 \%	&	18,2 \%	&	0 \%	&	0 \%	&	0 \%	&	0 \%	&	11	\\
\midrule
Unconscious bias 	&	0 \%	&	44,5 \%	&	22,2 \%	&	0 \%	&	0 \%	&	11,1 \%	&	22,2 \%	&	0 \%	&	0 \%	&	0 \%	&	9	\\
\midrule
Ethical software development / engineering ethics	&	20,0 \%	&	26,7 \%	&	26,7 \%	&	20,0 \%	&	6,60 \%	&	0 \%	&	0 \%	&	0 \%	&	0 \%	&	0 \%	&	15	\\
\midrule
Unprofessional behavior	&	0 \%	&	0 \%	&	66,7 \%	&	33,3 \%	&	0 \%	&	0 \%	&	0 \%	&	0 \%	&	0 \%	&	0 \%	&	3	\\
\midrule
Diversity in problem solving/product development/innovations 	&	35,7 \%	&	21,4 \%	&	21,4 \%	&	14,3 \%	&	0 \%	&	7,2 \%	&	0 \%	&	0 \%	&	0 \%	&	0 \%	&	14	\\
\midrule
Some other approach or example to the diversity and inclusion in SE	&	0 \%	&	33,4 \%	&	33,3 \%	&	0 \%	&	0 \%	&	0 \%	&	0 \%	&	33,3 \%	&	0 \%	&	0 \%	&	3	\\
\midrule
I don't use any examples about diversity \& inclusion in my course(s)	&	75,0 \%	&	0 \%	&	0 \%	&	0 \%	&	0 \%	&	0 \%	&	0 \%	&	0,0 \%	&	0 \%	&	25,0 \%	&	4	\\
\bottomrule
\end{tabularx}
\caption{Distribution of grading (Most used 1 --- 10 Least used) among respondents on different approaches and case examples of diversity and inclusion.}\label{tab:grading}
\end{table*}

In the structured question, "In your software engineering course(s), which case examples/approaches of diversity \& inclusion do you use the most in your teaching?" respondents were asked to rate various diversity and inclusion approaches and case examples on a scale of 1 to 10 (Most used=1, Least used=10). The respondents were required to grade at least one option, with an additional option to state that they do not use any examples about diversity and inclusion in their course(s). Respondents were guided not to give any grade for approaches they do not use, as the idea was not to grade all of the approaches but to investigate which approaches are used and which are the most used. Table~\ref{tab:grading} reports the overall distribution of the different approaches. 

Among the respondents, accessibility was the most familiar approach, with a total of 17 grades. Over half of the respondents reported that accessibility was the most commonly used approach or case example, making it the most frequently used approach compared to the others. The second and third most used approaches were diverse teams and diversity in problem-solving, with 15 and 14 gradings, respectively. Although ethical software engineering received 15 grades, it was not highlighted as one of the top three most commonly used examples, with only 20\% of respondents rating it as their number 1 choice.

Nine respondents graded unconscious bias, with 44\% rating it as the second most used approach for the diversity topic. Seven respondents graded the underrepresentation of women in software engineering and gender bias, and the grades were evenly distributed between grades 1 to 6. Only 3 respondents graded unprofessional behaviour, and it was rated as the third or fourth most commonly used approach.

Four respondents reported that they do not use any examples about diversity and inclusion in their course(s). Still, it could be assumed that one of these responses was misgraded with a 10 instead of a 1, as one respondent graded other diversity and inclusion approaches and that respondent has probably accidentally checked also the option "I don't use any examples about diversity and inclusion in my course(s)" among other options. 

Only three participants checked the option "Some other approaches to diversity and inclusion" in this question. However, in the subsequent open-ended question, "What other examples/approaches do you use in your teaching about diversity and inclusion?", a total of eight responses were provided. Notably, two of these responses stated that they did not use any other approaches besides those mentioned earlier. The remaining responses emphasized the importance of diverse teams in testing and, for example,  more comprehensive accessibility approaches, such as addressing colour blindness.

\subsection{Challenges observed in adding more diversity and inclusion topics to the teaching}

The challenges observed in adding more diversity and inclusion topics to the software engineering teaching were divided into thematic analysis to the challenges around perception of D\&I topics, time constraints, availability of resources, and need for allyship.

\paragraph*{Perception of D\&I Topics} In the open-ended responses, diversity and inclusion topics were often seen as ”soft” or ”humanistic” themes and were perceived as opposing technical topics. In these responses, respondents felt there could be challenges related to the perception of DEI as non-technical and, therefore, less relevant or valuable in a technical curriculum. Respondents were especially concerned about their students’ reactions toward adding more time and resources to the D\&I topics. Respondents worried that students would not expect to learn about D\&I topics but more technical skills. D\&I skills would be seen as outside the traditional scope of technical education.

\begin{quote}
    \textit{"Students are laser-focused on technology. Humanistic topics are not always welcome and have to be rationalized well."}
\end{quote}

\paragraph*{Time Constraints} Time troubled respondents in two ways. One was finding time to update courses to include more timely approaches to D\&I themes. Another aspect was the lack of time to identify which D\&I topics are most pressing and relevant to integrate into the rapidly evolving field of software engineering. In these responses, the respondents also found it hard to choose in which courses they could add more D\&I approaches and would need time to really sit on and plan the changes. 
\begin{quote}
    \textit{"Maybe the biggest issue is to take time to add inclusion topics to the course, maybe starting is the hardest - where and why to add more diversity themes to the course."}
	 \end{quote}
  
\paragraph*{Resource Availability} As many of the respondents were positive in their responses about adding more diversity and inclusion approaches to their courses, respondents also struggled to identify which diversity and inclusion topics were most timely and essential to be added to specific courses, as the curriculum in software engineering includes courses from various themes, and there could be various ways to approach D\&I themes. There was a lack of readily available resources that could be easily integrated into existing courses.: 
    \begin{quote}
\textit{"New information coming so fast, I have to find time to see if they are relevant for my courses"}
\end{quote}

\paragraph*{Need for allyship} If the respondents represented an underrepresented group, such as women, they were hoping to find allies, particularly from non-minority groups, to communicate and advocate for DEI topics effectively. It was felt that the message would be better delivered if the person talking about the subject would represent a non-minority, such as men. Respondents felt that men talking to men would be more impactful than women speaking to male students. 
\begin{quote}
    \textit{"Lack of accessibility resources as mentioned before, otherwise not really. I think some male students will always switch off if a woman is talking about gender issues, and that's typical of some young men, but none have ever been rude. I think that men talking about this would have more impact than me talking about it, though."}
\end{quote}

\subsection{Pedagogical strategies to approach diversity and inclusion in software engineering}

There were 15 respondents to the open-ended question, "\textit{Could you provide some examples of how and where you use diversity \& inclusion examples in your teaching}". However, in thematic analysis, it was discovered that different kinds of case stories and case examples were provided in all of the five open-ended questions. These cases were labelled as pedagogical strategies to approach diversity and inclusion in software engineering university-level education. Under the main theme \textit{Pedagogical Strategies for D\&I}, three different strategies were identified: a) Inclusive Curriculum Design, b) Active Learning and Co-creation and c) the Future of Software Engineering.

\paragraph{Inclusive Curriculum Design} Examples of incorporating D\&I in course content, such as case studies, history of women in computing, and teaching equitable design. Ensuring that slides, examples, and classroom communication are non-discriminatory and reflect diverse populations.

\textit{“I teach the difference between making things equal and equitable - equitable means some users will require additional support to achieve the same results as others. I ask groups to create personas and then highlight how most are a) Finnish b) white c) men and explain why this is a problem. I use examples from situations of male-led design that disadvantages women, e.g. emergency houses with no kitchen, male crash test dummies etc.”}

\paragraph{Active Learning and Co-creation} Methods that involve students in the learning process, like democratic decision-making and group formation strategies. Encouraging diversity in team compositions and using this as a learning point for the benefits of varied perspectives in problem-solving.

\textit{“My approach is to encourage students to create teams also with new people and not always with friends. If doing so, they meet new students and also see other ways of solving problems and working as part of the team. In this example, diversity is not forced or the only motivation for encouragement, but it is neutrally included in the process”}


\paragraph{Future of Software Engineering} Discussions around the broader impacts of software, including ethical considerations and social responsibility. Covering the implications of biased data in AI and the importance of critically examining data sources. Encouraging networking within diverse groups to prepare students for the realities of the tech industry.

\textit{“Teach and lead by example: the tiniest piece of example data must be diverse, and texts written in gender-free wordings, not assuming particular religion or ability. Then it scales up to all levels, including discussions of ethics and multicultural project work.”}

\section{Framework for D\&I in software engineering education}\label{sec:framework}

\begin{figure*}[ht]
\centering
\includegraphics[width=\textwidth]{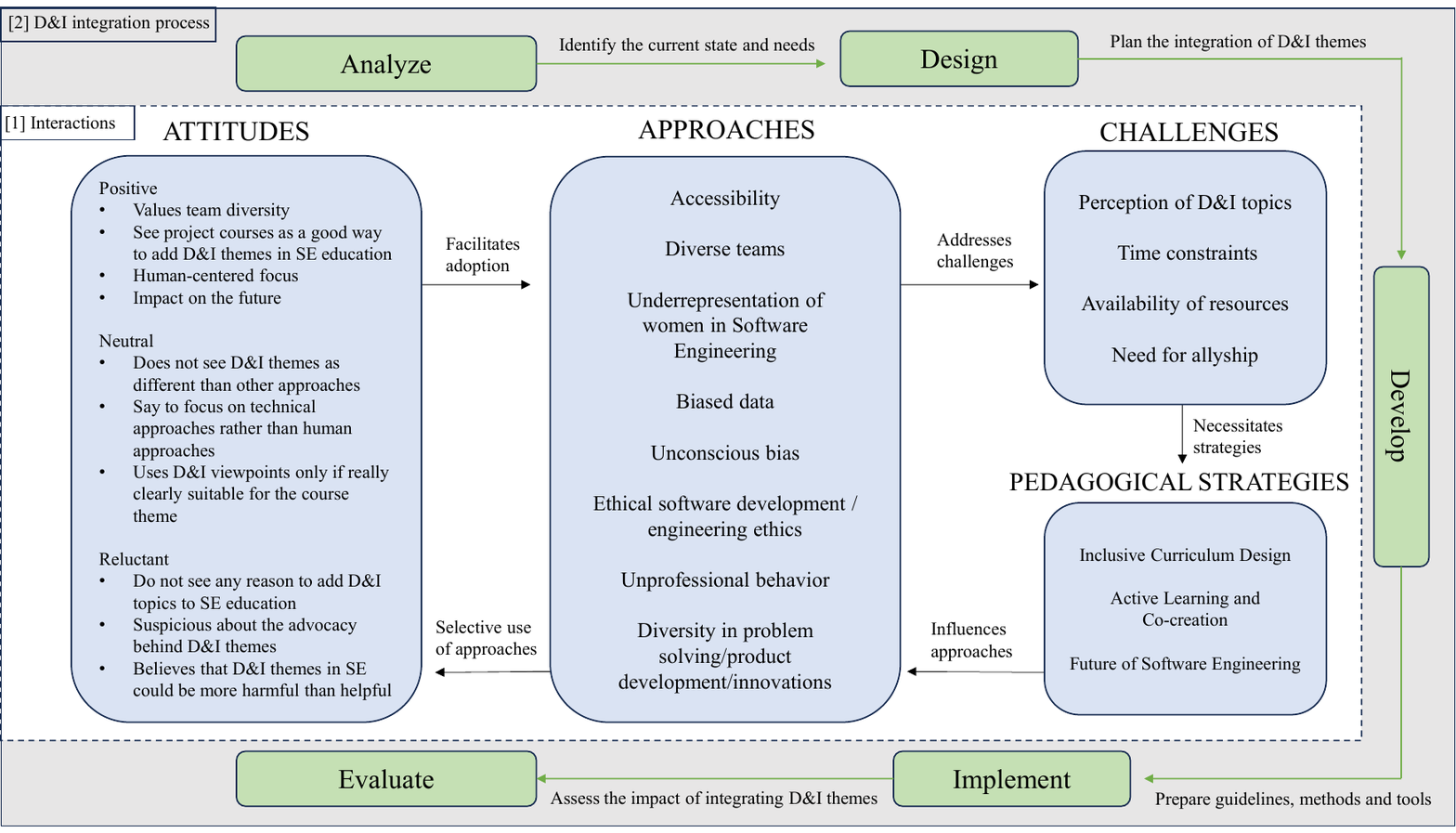} 
\caption{D\&I in SE Education Framework demonstrating the relationships between attitudes, approaches, challenges, and pedagogical strategies when discussing adding D\&I approaches to the university-level software engineering education, and steps for integrating D\&I in the teaching.}
\label{koonti}
\end{figure*}

This section presents a framework (Figure \ref{koonti}) designed to address the attitudes, approaches, challenges, pedagogical strategies and D\&I integration process steps identified in the thematic analysis of D\&I within software engineering education. Motivation and the need for the framework comes from the results of the overall satisfaction to the faculty’s diversity and inclusion actions presented in Subsection 4.1. Framework also visualizes the results of attitudes towards D\&I topics from Subsection 4.2., D\&I approaches used in software engineering education from Subsection 4.3., and challenges (Subsection 4.4.) and pedagogical strategies (Subsection 4.5.) identified from the data. Alongside visualizing the findings from the data, the framework also highlights the connections between different findings and gives recommendations for the steps for integrating D\&I topics for software engineering education.

This framework is formed for the software engineering faculty members to help them discuss the status of diversity and inclusion in their teaching and provide a systematic approach for integrating D\&I themes into software engineering education.

Framework has two dimensions: Interactions and the D\&I integration process: 
\paragraph{Interactions}
In this framework, 'Interactions' is for indicating the interaction and connection between different components of D\&I in software engineering university-level education:\\
\textbf{Attitudes}: Different attitudes facilitate the adoption of different approaches to the D\&I. There are three attitude viewpoints to the topic: Positive, Neutral, and Reluctant views towards D\&I themes.\\
\textbf{Approaches}: Key methods for integrating D\&I, such as accessibility or diverse teams. In some cases, some themes of D\&I were already naturally embedded in the course themes (such as accessibility), and that could have caused how teachers perceived D\&I topics and selected approaches.\\
\textbf{Challenges:} Issues that need to be addressed and be aware of when adopting different kinds of approaches, like perception of D\&I topics and time constraints.\\
\textbf{Pedagogical Strategies:} Inclusive Curriculum Design, Active Learning, and Future of Software Engineering. Used to solve challenges. More knowledge about different pedagogical strategies could also affect more different approaches implemented to the courses and, that way, positively affect the attitudes toward adding D\&I approaches for software engineering education. 

\paragraph{D\&I integration process}

The first step in the framework's second dimension, the D\&I integration process, is to analyze the current state and identify the needs related to D\&I themes in software engineering education. This involves understanding educators' attitudes towards D\&I. These attitudes also influence how D\&I approaches are adopted and how successfully they are integrated into the curriculum.

In the design phase, the focus is on planning how to integrate D\&I themes into the curriculum. This involves identifying key approaches already used or best practices and challenges faced in these approaches. 

The development phase involves creating the necessary guidelines, methods, and tools to implement the planned D\&I strategies. These are helped by different pedagogical strategies, which focus on solving the challenges and are meaningful for the education goals. 

The developed strategies, guidelines, approaches, and tools are put into practice in the implementation phase. This phase can involve actions such as providing access to the collected best practices or providing more resources for educators to modify their courses.

The final phase is to evaluate the impact of the integrated D\&I themes. This phase can involve actions such as collecting feedback from educators and students, identifying areas for improvement, and making necessary adjustments.

The D\&I integration process is systematic and iterative. After the evaluation phase, one can either continue collecting feedback or start the whole process from the analysis phase and move to designing and developing new guidelines and tools. During this process, transparent communication and involvement are important. However, this process dimension aims to give flexibility so that one can also use it to create one's courses towards a more diverse and inclusive approach and not necessarily do a whole faculty-level project.  

This framework aims to make the attitudes, approaches, challenges, pedagogical strategies and integrating process more visible when discussing D\&I in software engineering teaching. For example, the framework can be used when faculty members want to start a discussion or a process for integrating diversity and inclusion approaches into the curriculum. The framework emphasizes that there can be different viewpoints towards the topic and that there is a variety of approaches, which some of the faculty members can be already using in their courses (such as accessibility and biased data) and get in that way more ideas how to add more diversity and inclusion approach to them. Framework also helps to recognize different pedagogical strategies and how they can help solve the challenges when adding more diversity and inclusion approaches to software engineering education. Steps provided for the D\&I integration process help to identify at what stage faculty is in integrating D\&I topics in their curriculum and what steps they could take next. 

\section{Discussion}\label{sec:discussion}

\subsection{Key findings} 

The following subsection summarizes the survey's key findings regarding diversity and inclusion in software engineering teaching in Finland.

Firstly, the findings indicate that while some teachers hesitated to add diversity and inclusion topics to their courses, most respondents took a neutral or positive view in their responses and recognized the importance of diversity and inclusion in software engineering education.

Secondly, teachers were most familiar with accessibility approaches and often used them as an example of diversity and inclusion in their teaching. Diverse teams and the importance of diversity in problem-solving were also highly valued, particularly in project courses where teachers found it easy to incorporate team and communication skills. Furthermore, the justification for adding diversity and inclusion approaches near the technical mindset, such as biased data, was the easiest for teachers. 

Thirdly, challenges in adding diversity and inclusion topics to teaching were related to perceiving these topics as "soft" or unrelated to technical topics and a lack of time to update materials. Respondents also highlighted the difficulty in determining timely and essential diversity and inclusion topics for their courses. Minority respondents felt that having allies was important for discussing diversity and inclusion topics. Respondents used different kinds of pedagogical strategies to ease their possibilities of adding more D\&I approaches to the courses.

Fourthly, the respondents perceived increased diversity and multiculturalism in the academic field. However, there were signs of trust issues and confrontations between individuals who prioritize diversity and inclusion topics and those who do not, which may pose a challenge in promoting and implementing diversity and inclusion initiatives. Some respondents expressed pessimism about the reluctance of some teachers to engage with or incorporate relevant approaches into their courses as the primary obstacle to addressing diversity and inclusion issues.

Emerging from the results, a framework was formed for demonstrating the relationships between attitudes, approaches, challenges, and pedagogical strategies when implementing D\&I approaches to software engineering university-level education. This framework can guide software engineering educators to reflect on and improve their teaching strategies and walk through a structured D\&I integration process in their faculties. 

\subsection{Discussion}

In this study, support for incorporating diversity and inclusion themes in software engineering education was found, reflecting a growing recognition of the need for greater diversity in the field~\cite{forcecomputing}. This highlights the potential for such topics to be integrated more formally into the learning outcomes of university-level software engineering courses. For instance, several software engineering study programs in Finland use the ACM's Curricula Recommendations as a basis for their curricula. The results of this study also underline that best practices~\cite{blum2019computer, minerva-informatics-equality-award} from the computing discipline overall could be more utilized, as they could ease the adoption of D\&I in the teaching. 

A greater focus on diversity and inclusion themes in software engineering education could inform adjustments to the learning outcomes of relevant courses. For example, project courses such as Capstone courses~\cite{janzen2018reflection} and courses on data analysis could incorporate topics related to biased data. There are already positive signs that adding more equity, diversity, and inclusion approaches to the Capstone courses has its benefits~\cite{chromik2020teamwork}. The results of this study also support that it is possible to advance diversity and inclusion awareness via education as a part of bigger aim or learning outcome~\cite{hartwell2017breaking}. For example, while biased data were well covered in the responses, they were not always connected to an overall understanding of diversity and inclusiveness.

The development and training of faculty members are crucial for successfully implementing a more diverse-aware approach in various fields, including software engineering~\cite{sciame2009infusing}. As this research shows, the process of incorporating diversity and inclusion into software engineering education is ongoing, and attitudes toward these topics may depend on the overall climate of the faculty. Some respondents in this study still viewed technical skills as the sole important factor, echoing similar discussions in software development team research~\cite{kohl2018perceptions}. Different academic disciplines are in different stages of adding more D\&I to their curriculum. However, this study reveals that the field of software engineering is not among the most advanced in this regard. The pedagogical strategies found in this research would implicate that there would be a need for more concrete guidelines and actions to increase the visibility of D\&I topics in software engineering education. For that need, the framework in this study was formed.

Experiences from other fields can already give us concrete tools and strategies to implement D\&I topics in teaching~\cite{hartwell2017breaking}. For instance, medical education has developed comprehensive frameworks for embedding cultural diversity and inclusion into their curricula, emphasizing institutional policies, curriculum development, faculty training, and assessment~\cite{dogra2009twelve}. These frameworks could serve as models for software engineering education, especially helping to understand cultural or power differences in software engineering. This can be more timely, as we are entering the era of Generative AI, and that is changing software engineering but also how software engineering impacts the world~\cite{hyrynsalmi2024bridging}.

The results of this research suggest avenues for future research, encouraging us to continue data collection and supporting teachers in identifying how to incorporate more diverse and inclusive approaches in their courses. There is a particular need for best practices in deeply technical courses, and it is important to train future professionals with a strong background in diverse and inclusive software engineering and development. The framework for D\&I in software engineering education is created to help faculties, and individuals approach this complex topic, embrace the variety of attitudes and get the structured process to integrate D\&I topics for teaching. 

Overall, this study highlights the continuing need for diverse-aware education and training and emphasizes the responsibility of universities to ensure that future professionals are equipped with the necessary skills and knowledge to promote diversity and inclusion in software engineering.

\subsection{Limitations and future study} 

The first and most visible limitation of this study is its narrow focus on Finnish universities, which may limit the generalizability of its results to other cultural and institutional contexts. In addition, Finland's university-level engineering education is concentrated on a handful of well-connected universities, and the local setting might bias the results. However, given that the data was collected via an online survey in English and did not have any Finnish-specific questions, it remains possible to expand the study's scope and collect data from a broader range of participants globally. In that phase, the question about the academic position should also be added to add a more comprehensive understanding of the data and its applicability across different academic levels and framework conducted in this research test.

Secondly, despite authors best efforts to attract respondents unfamiliar with D\&I themes, often those with strong opinions for or against the surveyed topic are answering. A case study on a particular selected university could be carried out in future inquiries to countermeasure this bias. This kind of approach, with interviews and more personalized connections, would also shed light on those teachers who would not answer this kind of survey.

Thirdly, the study acknowledges the inherent complexity of the terms "diversity" and "inclusion". To address this issue, the researchers made efforts to ensure that respondents had a shared understanding of these terms by providing explanations and usage examples at various points throughout the survey. Despite these efforts, it is possible that there were still some misunderstandings around terms.

To address this potential bias, future research could consider incorporating questions about respondents' level of familiarity and confidence with diversity and inclusion topics. It may also be beneficial to explore whether these topics could be framed in a more familiar context for software engineers, such as soft skills while ensuring they are effectively identified and measured.


\section{Conclusions}\label{sec:conclusions}

The attitudes and perceptions of software engineering educators towards D\&I are critical in developing effective strategies to promote these concepts in education and faculty. In this study, it is investigated how D\&I issues are approached in software engineering education and what kind of perceptions and challenges there are. Findings of the study suggest that while some educators are neutral or reluctant to include diversity and inclusion topics in their courses, positive respondents recognize their importance, particularly in accessibility approaches, diverse teams, and the importance of diversity in problem-solving. However, incorporating D\&I topics into teaching presents challenges, including how to start the process in the faculty level, understanding what would the most important approaches for the topic and the lack of time to modify courses. Additionally, trust issues and confrontations were observed between individuals who prioritize D\&I topics and those who do not, suggesting a potential challenge in promoting and implementing diversity and inclusion initiatives. As a result, a framework for integrating D\&I and demonstrating the relationships between attitudes, approaches, challenges, and pedagogical strategies to add D\&I approaches to software engineering education is presented. This framework provides valuable insights into diversity and inclusion in software engineering education in Finland and beyond.

\section{Acknowledgements}
Funding: This research was supported by PHP Säätiö (PHP Holding), Grant 20230012. I want to thank the anonymous reviewers and colleagues who have provided valuable comments and feedback on this research and the framework.


\end{document}